# Ultrasonic and Electromagnetic Sensors for Downhole Reservoir Characterization

K. Wang, H. T. Chien, S. Liao, L. P. Yuan, S. H. Sheen, S. Bakhtiari, and A. C. Raptis

Argonne National Laboratory
9700 South Cass Avenue
Argonne, IL 60439
Email: kwang@anl.gov

## ABSTRACT

The current work covers the evaluation of ultrasonic and electromagnetic (EM) techniques applied to temperature measurement and flow characterization for Enhanced Geothermal System (EGS). We have evaluated both ultrasonic techniques and microwave radiometry for temperature gradient and profile measurements. A waveguide-based ultrasonic probe was developed to measure the temperature gradient. A statistic approach on estimating the average grain size via spectral analysis of the scattered ultrasonic signals is introduced. For directional temperature measurement, different microwave antenna designs are compared numerically and an array loop antenna design is selected for further development. Finally techniques to characterize the porosity and permeability of a hot dry rock resource are presented.

## INTRODUCTIONS

Geothermal energy is an established form of alternative energy that is being harvested in many locations around the world. An almost limitless supply of energy is available from the core of the Earth, arising as hot spots near the surface, which can be utilized by geothermal power plants. It is estimated that the total heat flow over the surface of the Earth is approximately $4.4 \times 10^{13}$ J/s, mainly as a result of the planet cooling and the decay products of radioactive rock (Gupta and Roy, 2007). The traditional geothermal energy process is limited to natural convective hydrothermal sources, where water and porous rock are present in the necessary quantities. The water passes through the porous rock, absorbing the heat from the rock and then is pumped to an above ground steam electricity generator. The availability of water resource restricts this traditional process. The Enhanced Geothermal System (EGS) creates artificial geothermal reservoirs and thus increases possible locations for geothermal power (Tester, et al. 2007 and 2008). A well is drilled into a suitable site and then developed based on hydraulic stimulation (Fokker, 2006). The process uses pressurized water to fracture the rock, creating micro-fractures and porous surfaces, which increase the surface area for heat transfer. The water flows through the cracks and absorbs the heat. The resulting hot water/steam mixture is then pumped to the surface and into the power generator. Studies have estimated that for the fluid entering the generator, at 200°C, a mass flow rate of 80 kg/s is needed to attain economic viability for the reservoir (DOE-EERE, 2008).

A successful EGS needs reliable instrumentation to control the logging process and monitor the geothermal well/reservoir properties. Sensors to measure key EGS reservoir parameters, including directional temperature, pressure, fluid flow, and flow/rock interaction, are essential to the system performance. At present, EGS reservoir performance, as derived from reservoir geometry and permeability, contains significant uncertainties. In particular, the reservoir temperature gradients and flow rates of circulating fluids in an EGS reservoir are major unknowns. The current downhole logging tools and sensors are mainly developed for gas and oil exploration, applied to an environment of relatively low temperature (Timur and Toksoz, 1985; Qi et al., 2002). The reservoir temperature of a geothermal well is typically higher than 200°C, which limits direct applications of the current gas/oil well instruments that can last only a brief period with heat shielding. In addition, oil/gas wells are located in predominately sedimentary formations while geothermal wells are more often located in hard crystalline rocks normally of volcanic origin. Therefore, high-temperature logging sensors for measuring temperature, pressure, flow, lithology, and other reservoir characteristics are needed in order to successfully develop and economically operate an EGS reservoir (Sheen et al. 2010).

With considerable experiences on developing real-time ultrasonic inspection techniques for hostile environment application (e.g. defect inspection and component identification under molten sodium) (Sheen et al. 2009, Chien et al. 2010, Wang et al. 2010, and Sheen et al. 2010), we have conducted laboratory study on developing 1) A waveguide based temperature probe for temperature gradient measurement; 2) A spectral analysis method to estimate the average grain size statistically; and 3) A microwave radiometer for directional temperature measurement.

**TEMPERATURE PROBE**

An ultrasonic temperature probe that potentially can be applied to downhole temperature profile measurement in the enhanced geothermal systems uses a magnetostrictive (MS) transducer to generate surface guided waves in a low-loss magnetostrictive alloy rod (waveguide).

A couplant is always required for an ultrasonic measurement using traditional transducer, but MS transducers acts through totally different physical principles (Thompson, 1990). When an electrically conducting object (metal wire or plate) is placed at the center or near the surface of an MS transducer, eddy currents will be induced in a near surface of the object. With the presence of a static magnetic field, the induced eddy currents will be subjected to Lorentz forces, which create surface guided waves. The most important advantage of MS transducer is couplant free transduction, which allows operation without contact at elevated temperatures and in remote locations, and produces highly consistent measurements.

Surface guided waves have been widely exploited for the non-destructive evaluation (NDE) of large and complex structures. One the most attractive properties is that, surface guided waves are able to propagate significant distances in either pitch-catch or pulse-echo modes, which enable it to accumulate information along their propagation path. Furthermore, it has the capability to follow the curvature of the waveguide and can be directed along the desired path.

This temperature probe has a simple design, which only consists of an MS transducer (Panametrics ET100) and a low-loss magnetostrictive alloy rod ($\Phi = 0.03$"). The probe can be bent or embedded with notches to facilitate temperature profile measurements. As surface guided waves propagating along the rod, the times-of-flights (TOF) of reflections from each bend or notch are analyzed. These TOFs are directly related to the temperature between the bending points or notches. Laboratory tests were conducted in mineral oil up to 260°C. Figure 1 illustrates the schematic drawing of the temperature probe and its performance at elevated temperatures. Two reflections, from the bending point and the end tip of probe, are detected. As the oil temperature rising, the travel times of the reflections increase accordingly, but the shift depends on the temperature profile of the probe. Higher temperature would induce bigger shift on TOF.

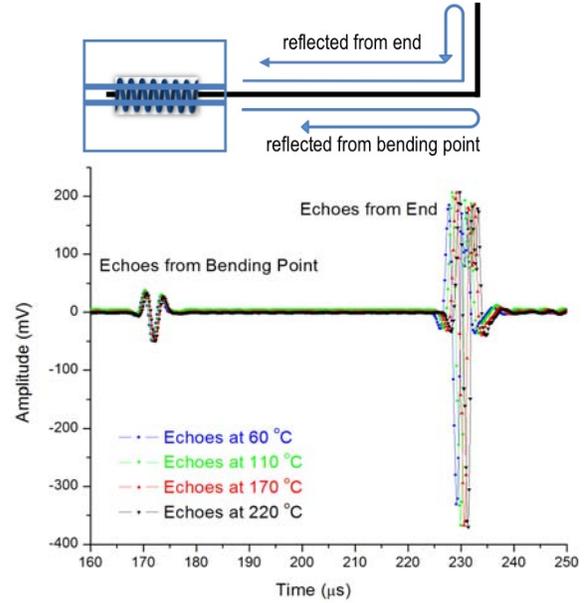

*Figure 1: Reflections from a bent probe for different temperatures.*

A close look on the TOF expression of the echoes will reveal the factors that determine the shift. The TOF of each echo has a simple expression:

$$t = \frac{2l}{V} \qquad (1)$$

Where $l$ is the travel distance of the signal and $V$ is the sound velocity in the waveguide. Then the difference of TOF under difference temperatures can be written as:

$$\delta t = \frac{2\delta l}{V_0^2} - t_0 \cdot \frac{\delta V}{V_0} \qquad (2)$$

Where

$$\delta t = t_{T_0} - t_T, \ \delta l = l_0 \alpha (T_0 - T),$$
$$t_0 = \frac{2l_0}{V_0}, \quad \delta V = V_0 - V_T, \qquad (3)$$
$$V_0 = V(T_0), \ V_T = V(T),$$

and $\alpha$ is the linear thermal expansion coefficient of the waveguide. For the chosen alloy, its $\alpha$ is a constant and $V(T)$ is linearly decreasing within the tested temperature regime. Consequently, $\delta t$ at the bending point and the end tip can be accurately measured. A linear relationship between $\delta t$ and $T$ is expected.

Figure 2 shows the travel time changes, $\delta t$, of the reflected echoes as the surrounding temperature increases. Linear relationship between $\delta t$ and $T$ are confirmed for the measurements from both the bending point and end tip. A linear least squares regression was employed and the results are shown in Figure 2. The corresponding coefficients of determination $R^2$ are also provided, which provide a measure of how well higher temperature measurements are likely to be predicted by the fitted model. By monitoring the difference between two fittings, we are able to measure the temperature inversely. The error of the prediction is within ±7%. With more notches along the probe, we are able to map the temperature gradient by measuring the time interval between each pair of the reflected echo and comparing with the fitted model.

Figure 3 shows the pressure effect on an ultrasonic temperature probe. Over a wide range of pressure, the variation of TOF is insignificant.

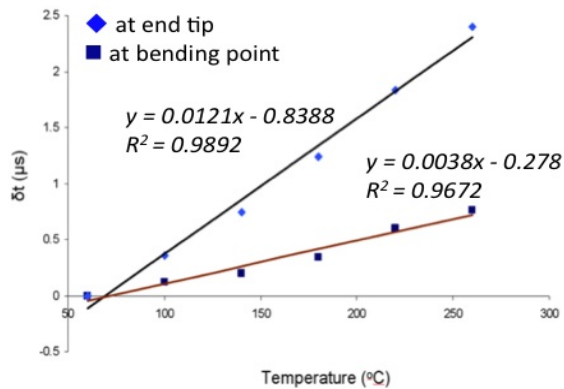

*Figure 2: Time-of-flight changes as a function of oil temperature.*

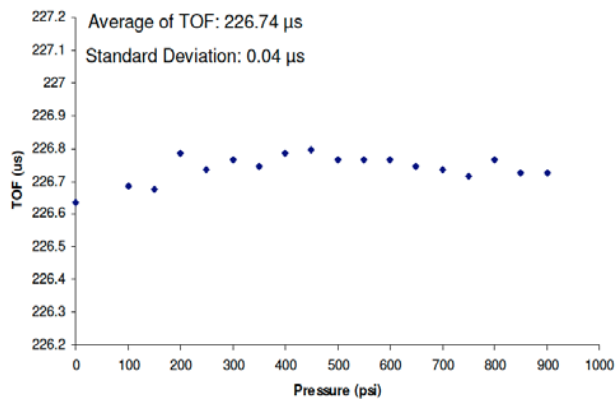

*Figure 3: Effects of pressure on the ultrasonic temperature probe.*

## SPECTRAL ANALYSIS OF SCATTERED ULTRASONCI SIGNALS

Grain size estimation in polycrystalline materials using scattered grain boundary echoes has significant practical applications in determining the quality and structural integrity of materials (Aduda and Rawlings, 1996). It has been successfully deployed as a NDE method, since the average grain size can help to identify certain material characteristics. However this method has been mainly used for homogeneous solid materials (metals, glass, and crystals minerals). Because of the complexity of ultrasonic waves propagating and scattering in inhomogeneous materials, therefore, its application in geomaterials (rocks) has not been well studied. In this work, we introduce an alternative approach, spectral analysis, to study the ultrasonic scattering and its relation to grain boundary, grain size, grain orientation, and composition of rocks.

*Table 1: Compositions and pictures of the rock samples used in the experiment.*

| 1. Gneiss | |
|---|---|
| | Mineralogy: |
| | 60% brown and white feldspar(1-5 mm) |
| | 30% quartz grain (1-4 mm) |
| | 5% biotite (1-2 mm) |
| | 5% garnet (1-4 mm) |
| | Texture: Foliated |
| 2. Charnockite (orthopyroxene-bearing granite) | |
| | Mineralogy: |
| | 40% translucent feldspar (up to 20 mm) |
| | 30% brown quartz (5-10 mm) |
| | 30% orthopyroxne |
| | Texture: Coarse grained |
| 3. Gabbro (Norite) | |
| | Mineralogy: |
| | 50% pyroxene |
| | 40% plasgioclase |
| | 10% biotite |
| | Texture: Fine grain (1-3mm) |

While propagating in rocks, ultrasonic waves are subject to scattering and mode conversions and are dispersive because the inhomogeneity of the rocks. The scattering signals may be non-explicitly related to rock's average grain size and grain orientation. Therefore, it is feasible to statistically estimate the average grain size with proper analysis on the scattered echoes. Our method focuses on analyzing the spectral signatures of different rocks to determine grain size and formation.

A laboratory study on the ultrasonic scatterings of three rock samples was conducted. Detail geological information of the samples is given in Table 1. The experimental set-up consists of a signal generator and receiver (Panametrics 5058PR), a LeCroy oscilloscope (LeCroy 9370), and two rectangular transducers (Panametrics A413S, 0.5 MHz, 0.5"×1"). The transducers are mounted on a bronze chase with a fixed distance of 3" (center to center). In the experiment, the measurements were taken under pitch-catch mode, and the transmitter was excited at 100 V pulse voltage and with 40 dB signal gain. For each sample, ultrasonic scatterings at various locations were measured to acquire representative results. Each received signal was averaged for 200 iterations to improve the signal to noise ratio (SNR). Figure 4. shows the received scattered ultrasonic signals and their FFT.

A close look on the signals of sample 1, 2, and 3 indicates that, the received echo of sample 3 contains more high-frequency components, and may have a broader spectrum in frequency domain than the other two samples, which is confirmed by the FFT results. The FFT results further indicate an inverse relationship between the average grain size and the peak position in FFT spectra. For quantitative results, further calculation on the power spectral density (PSD) is required, which is able to provide the power distribution of a signal in the frequency domain. The relationship between PSD and the average grain size ( ) and standard deviation ($\sigma$) can be written as (Bilgutay et al. 1989):

$$\Phi(\omega) = \phi^2 \frac{[1 - 2\exp(-\frac{2N\sigma^2}{c^2}\omega^2)\cos(\frac{2N\phi}{c}\omega) + \exp(-\frac{4N\sigma^2}{c^2}\omega^2)]\exp(-\frac{N\sigma^2}{c^2}\omega^2)}{1 - 2\exp(-\frac{2\sigma^2}{c^2}\omega^2)\cos(\frac{2\phi}{c}\omega) + \exp(-\frac{4\sigma^2}{c^2}\omega^2)}$$

(4)

Where $c$ is the sound velocity in the sample, $\omega$ is the angular frequency and $\omega = 2\pi f$. A PSD is a function of frequency ($\omega$), average grain size ( ), standard deviation ($\sigma$), and the number of grains ($N$) along the path. For an ideal case, $\sigma$ is infinitesimal; Equation (4) then is simplified as:

$$\Phi(\omega)_{\sigma \to 0} = \phi^2 [\frac{\sin(\frac{N\sigma}{c}\omega)}{\sin(\frac{\sigma}{c}\omega)}]^2 \qquad (5)$$

Using the 6 dB frequency spacing or bandwidth of the peak $\Delta f$, the expression of the average grain size ($\sigma$) is further simplified as:

$$\Delta f = \frac{c}{2\phi} \qquad (6)$$

For practical cases, $\sigma$ is always non-zero. Quantitative studies on estimating the average grain size and the standard deviation using spectral analysis are in progress and results will be reported in future studies.

**MICROWAVE RADIOMETER**

Microwave radiometry is a well-established passive technique for remote temperature sensing (Janssen, 1993; Stephan, 2005 and 2006). With narrow-band filtering, it can also provide chemical analysis. The basic principle of microwave radiometry is based on Planck's black body radiation from the probed materials. In drilling a geothermal well, it is desirable to know the right direction, the direction towards the hot reservoir. To date, such a temperature sensor for in-situ reservoir temperature profile measurement is not readily available. We are conducting a feasibility study to evaluate the microwave radiometry technique for downhole applications. The key challenge of this technique is the antenna design that can sustain the harsh environment and provide remote sensing of reservoir temperature with desired sensitivity and directivity.

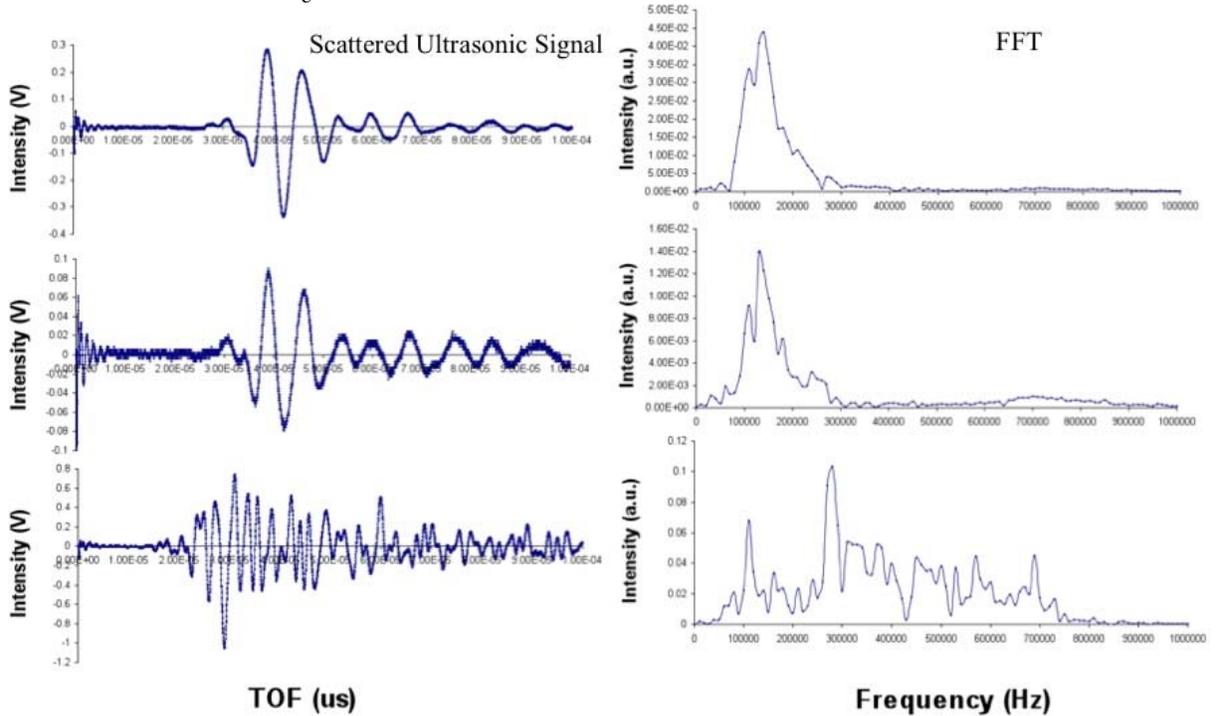

*Figure 4: Scattered Ultrasonic Signals and FFT.*

## Fundamentals

A microwave radiometer is a passive temperature sensor that detects electromagnetic radiation from the probed object. The power radiated from the object at temperature $T$ follows the Planck's black body formula that at the Rayleigh-Jeans limit has the simple expression (Janssen, 1993),

$$B_v(T) \approx \frac{2kT}{\lambda^2} \quad (7)$$

Where $B_v(T)$ is the power per unit area, $k$ is the Boltzmann constant, and $\lambda$ is the wavelength. The black body radiation experiences attenuation due to both free space attenuation and absorption in the propagating medium, which can be described by the Friis formula (Janssen, 1993),

$$P_v(T) = A_e A_{source} B_v(T) \frac{\exp(-r/l)}{4\pi r^2} \quad (8)$$

Where $A_{source}$ and $A_e$ are the areas of heat source and effective antenna, respectively, $\ell$ is the absorption length in the medium, and $r$ is the distance of antenna from the source. The effective antenna area $A_e$ is linearly dependent on the gain, $G$,

$$A_e = \frac{\lambda^2}{4\pi} G \quad (9)$$

Substituting Eq. (7) and Eq. (9) into Eq. (8) gives

$$P_v(T) = kTGA_{source} \frac{\exp(-r/l)}{8\pi^2 r^2} \quad (10)$$

Eq. (10) simply says that the detected power doesn't depend on frequency rather it depends linearly on the antenna gain $G$. Including the antenna bandwidth, $B$, the total detected power is given by

$$P(T) = BP_v(T) \quad (11)$$

To estimate the detection sensitivity of a microwave radiometer depends on the total noise temperature $T_N$ the antenna bandwidth B and the total integral time $\tau$. At $T_N$, the noise level, $N$ can be estimated from

$$N(T_N) = kT_N \sqrt{\frac{B}{\tau}} \quad (12)$$

The ratio of $P(T)$ and $N(T_N)$ represents the signal-to-noise ratio (SNR) given as

$$SNR = GA\sqrt{B\tau} \frac{T\exp(-r/l)}{8T_N \pi^2 r^2} \quad (13)$$

From Eq. (13), we see that for a given temperature $T$ and noise temperature $T_N$ of the object, the SNR is linearly proportional to the antenna gain and square-root dependent on the antenna bandwidth. By setting SNR=1, the expected detection sensitivity (minimum detectable temperature) of the microwave radiometer is given by

$$T_{min} = \frac{8\pi^2 r^2 T_N}{GA_{source}\sqrt{B\tau}\exp(-r/l)} \quad (14)$$

Based on Eq. (14) a broadband high-gain antenna with a large effective antenna area will be the optimal choice. The desirable antenna choice is a broadband high-gain helix antenna that operates in a frequency range from 460 MHz to 900 MHz (B = 440 MHz). Such an antenna working in axial mode has a gain of 25 (14 dBi). With $r$ = 10 m, integrating time $\tau$ = 1 s, and $T_N$ = 2500K, the estimated sensitivity calculated by Eq. (14) gives $T_{min}$ = 50K.

## Antenna Selection

Table 2. lists characteristics of three viable antenna designs, in which helix and conical spiral antennae are single antenna design. To reduce the horizontal antenna size for realistic downhole applications, we evaluate the array loop antenna design. A small loop antenna has a gain of

$$G_{loop}(\theta) \propto |\sin(\theta)|^2 \quad (15)$$

Where $\theta$ is the beam diversion angle. The benefit of the loop antenna is its small size but the antenna gain is reduced. However, an array of loop antenna can be used to increase the gain. The total radiation pattern $G(\theta)$ of an array loop antenna is the product of the single loop antenna radiation pattern, $G_{loop}(\theta)$ and an array factor $AF(\theta)$, given as

$$G(\theta) \propto G_{loop}(\theta) \times |AF(\theta)|^2 \quad (16)$$

Giving an $N$-element array with spacing of $d$, the array factor can be given as

$$AF(\theta) = A_0 \sum_{n=0}^{N-1} e^{jknd\cos\theta} = A_0 \frac{\sin(\frac{N}{2}kd\cos\theta)}{\sin(\frac{1}{2}kd\cos\theta)} e^{j\frac{N}{2}kd\cos\theta} \quad (17)$$

Where $k$ is the wave number and $A_0$ is a constant. $AF(\theta) = NA_0$ when $\theta = 90°$. Combining Eqs. (15) and (17), the total array gain pattern is given by

$$G(\theta) = \left|\sin\theta \frac{\sin(\frac{N}{2}kd\cos\theta)}{\sin(\frac{1}{2}kd\cos\theta)}\right|^2 \quad (18)$$

Figure 5 shows the gain changes as a function of number of loops, $N$, for the design of 100 MHz microwave with half wavelength in spacing, $d$. It is clearly shown that at 90° the beam gain is directly proportional to $N$. Due to the horizontal size constraint for downhole application, the array loop antenna design is chosen for further development.

*Table 2: Characteristics of different antenna designs*

| Antenna Type | Helix Antenna | Conical Spiral Antenna | Loop Array (N=4 to 10) |
|---|---|---|---|
| Image | 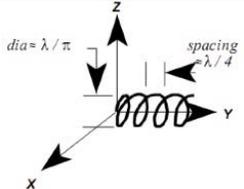 | 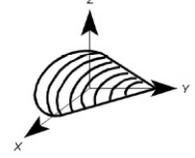 | 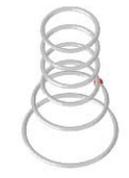 |
| Radiation Pattern | Evaluation & Azimuth 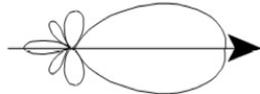 | Evaluation & Azimuth 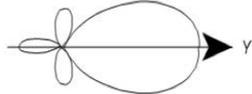 | Evaluation 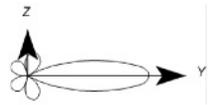 Azimuth 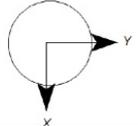 |
| Workning Frequency $F_C$ in Air | 100 MHz – 3 GHz | 50 MHz – 18 GHz | Arbitrary |
| Workning Frequency $F_C$ in Water | 11 MHz – 31 MHz | 6 MHz – 2 GHz | Arbitrary |
| Gain | 10 dB – 20 dB | 5 dB – 8 dB | 4 dB – 10 dB |
| Bandwidth | 52% of $F_C$ | 120% of $F_C$ | Arbitrary |
| Beam Width | $50^O – 75^O$ | $60^O – 80^O$ | $20^O – 60^O$ |
| Typical Size | **Horizontal size** 1-2 Wavelength $\lambda_C$ (33 cm – 66 cm) **Vertical size** 2-5 Wavelength $\lambda_C$ (0.7 m – 1.7 m) $F_C$ = 10 MHz | **Horizontal size** < 2 Wavelength $\lambda_C$ (< 66 cm) **Vertical size** 2-5 Wavelength $\lambda_C$ (0.7 m – 1.7 m) $F_C$ = 10 MHz | **Horizontal size** 0.5-1 Wavelength $\lambda_C$ (17 cm – 33 cm) **Vertical size** 2-5 Wavelength $\lambda_C$ (0.7 m – 7 m) $F_C$ = 10 MHz |
| Sensitivity | 50 K – 100 K | 60 K – 120 K | 60 K – 120 K |

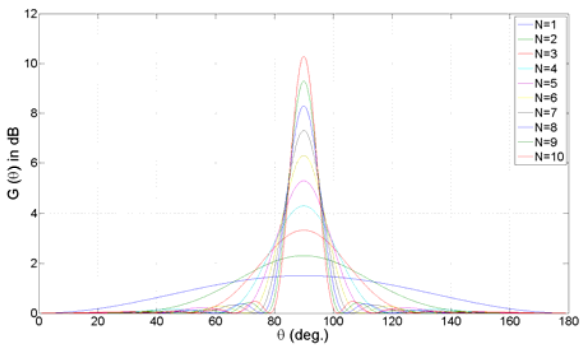

*Figure 5: Array gain pattern, G(θ), for N=1 to 10.*

### CONCLUSIONS AND FUTURE WORKS

For a better understanding of a geothermal reservoir, especially EGS, the essential parameters that need to be determined are temperature gradient, rock formation, porosity and permeability, and flow rate. To date, limited measurement techniques have been developed for reservoir characterization and downhole sensors and instrumentation for in-situ measurement are scarce. In the current work, we have conducted a laboratory study on ultrasonic and EM techniques applied to temperature measurement and rock formation characterization. For temperature measurement, we have evaluated the microwave radiometery and ultrasonic techniques for temperature gradient and directional temperature measurements. An ultrasonic temperature probe based on EMAT was constructed and assessed. Antenna design for a downhole microwave radiometer was analyzed; an array loop-antenna design was selected for further development. For study of grain size of rocks, spectral analysis of the ultrasonic scatterings was conducted. The FFT results reveal an inverse relationship between the average grain size and the major peak locations in analyzed spectra.

Permeability and connectivity of the reservoir are also critical parameters, especially for hot dry rock (HDR) resources. The efficiency of energy extraction from HDR reservoir is directly related to them. A feasibility study by monitoring the flow induced noise level from a hot rock test facility will be carried out in the future, which would also provide useful information on the development of hydraulic stimulated fracture during the operation.

*Acknowledgement*

This project is supported by the US. Department of Energy, Assistant Secretary for Energy Efficiency and Renewable Energy, under DOE Grant No. DE-PS36-09GO99017.